\documentclass[aps,prl,reprint,showpacs]{revtex4-1}

\usepackage{graphicx}
\usepackage{subfigure}
\usepackage{bbm}
\usepackage[pdftitle={Experimental protocol for high-fidelity heralded photon-to-atom quantum state transfer},pdfauthor={Christoph Kurz},pdfdisplaydoctitle=false,bookmarks=false,pdflang={en-US}]{hyperref}
\newcommand{\ket}[1]{\left|#1\right\rangle}
\newcommand{\SState}{\text{S}_{1/2}}
\newcommand{\PState}{\text{P}_{3/2}}
\newcommand{\DState}{\text{D}_{5/2}}
\newcommand{\pState}{\ket{+\frac{1}{2}}}
\newcommand{\mState}{\ket{-\frac{1}{2}}}
\newcommand{\pmState}{\ket{\pm\frac{1}{2}}}
\newcommand{\pSigma}{\ket{\sigma^+}}
\newcommand{\mSigma}{\ket{\sigma^-}}

\newcommand{\ppi}{\ket{\pi}}
\newcommand{\Ca}{^{40}\text{Ca}^+}

\begin{document}

\title{Experimental protocol for high-fidelity heralded\\photon-to-atom quantum state transfer}

\author{Christoph Kurz}
\email{c.kurz@physik.uni-saarland.de}
\author{Michael Schug}
\author{Pascal Eich}
\author{Jan Huwer}
\author{Philipp M\"uller}
\author{J\"urgen Eschner}
\affiliation{Universit\"at des Saarlandes, Experimentalphysik, Campus E2 6, 66123 Saarbr\"ucken, Germany}


\maketitle

\textbf{A quantum network combines the benefits of quantum systems regarding secure information transmission and calculational speed-up by employing quantum coherence and entanglement to store, transmit, and process information. A promising platform for implementing such a network are atom-based quantum memories and processors, interconnected by photonic quantum channels. A crucial building block in this scenario is the conversion of quantum states between single photons and single atoms through controlled emission and absorption. Here we present an experimental protocol for photon-to-atom quantum state conversion, whereby the polarization state of an absorbed photon is mapped onto the spin state of a single absorbing atom with $>$95\,\% fidelity, while successful conversion is heralded by a single emitted photon. Heralded high-fidelity conversion without affecting the converted state is a main experimental challenge, in order to make the transferred information reliably available for further operations. We record $>$80\,s$^{-1}$ successful state transfer events out of 18,000\,s$^{-1}$ repetitions.}

Trapped single atomic ions are well suited for implementing quantum memories and processors, as they allow for long quantum information storage times and high-fidelity state manipulation and read-out \cite{Leibfried2003, Langer2005, Benhelm2008, Schindler2013}. Optical photons, on the other hand, are the most viable carriers of quantum information over long distances \cite{Weihs1998, Ma2012}. The integration of ion-based quantum processing nodes and photonic quantum communication in order to form a quantum network \cite{Kimble2008, Duan2010} appears therefore particularly attractive. Remarkable fidelity and efficiency have been achieved in experiments on photon storage and read-out with atomic ensembles \cite{Tanji2009}; we focus on single-atom systems because of their prospect of processing the transmitted quantum information via local quantum gate operations \cite{Hucul2014}. Using single ions or atoms as senders and receivers of single photons requires controlling the emission \cite{Stute2013} and absorption processes, which, however, suffers from low efficiencies, most of all because of incomplete overlap between the spatial profile of the atomic emission or absorption and the photonic mode \cite{Schug2013}. Even with approaches that employ optical resonators \cite{Boozer2007, Specht2011, Ritter2012} or a deep parabolic mirror \cite{Maiwald2009, Fischer2014}, results have so far remained limited to overall fidelities below 10\,\%. A possible remedy, which makes the fidelity of the state transfer process independent from its efficiency, is to filter out successful events by a heralding signal; this has been proposed \cite{Cabrillo1999, Feng2003, Simon2003} and successfully applied for the creation of remote atom-atom entanglement \cite{Moehring2007, Hofmann2012}. It has also been proposed for mapping a photonic polarization state onto an atomic quantum bit by heralding the single-photon absorption with the detection of another single photon emitted in a Raman process \cite{Lloyd2001, Mueller2014}, but experimental approaches so far have only used a fluorescence signal as herald, which does not preserve the atomic state \cite{Piro2011, Huwer2013, Schug2013}.

Here we present the experimental implementation of a protocol for high-fidelity photon-to-atom state mapping heralded by Raman emission of a single photon. Out of 18,000\,s$^{-1}$ runs of the protocol, we obtain $>$80\,s$^{-1}$ successful state transfer events with 97\,\% state fidelity and 95\,\% process fidelity. The single absorption events in this proof-of-principle experiment are driven by multi-photon laser pulses, but the results demonstrate that the state transfer protocol may be readily implemented with true single photons.

\subsection*{Results}

\begin{figure}[ht]
   \renewcommand{\thesubfigure}{\hspace{0.09\columnwidth}\smash{(\alph{subfigure})}}
   \subfigure[]{\includegraphics[height=0.29\columnwidth]{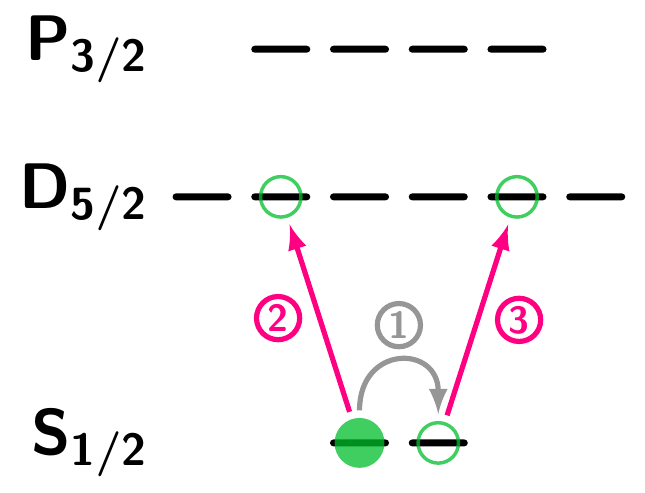}}
   \hfill
	\renewcommand{\thesubfigure}{\smash{(\alph{subfigure})}}%
   \subfigure[]{\includegraphics[height=0.29\columnwidth]{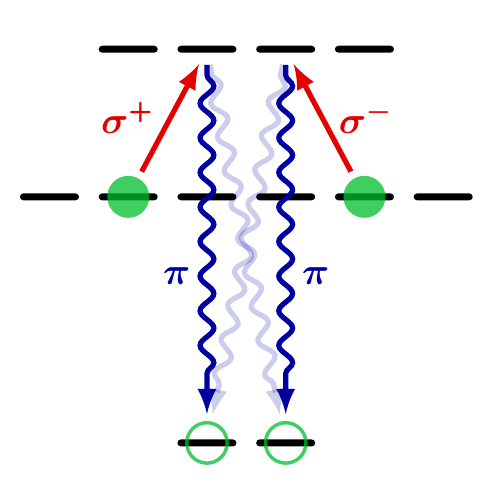}}
   \hfill
   \subfigure[]{\includegraphics[height=0.29\columnwidth]{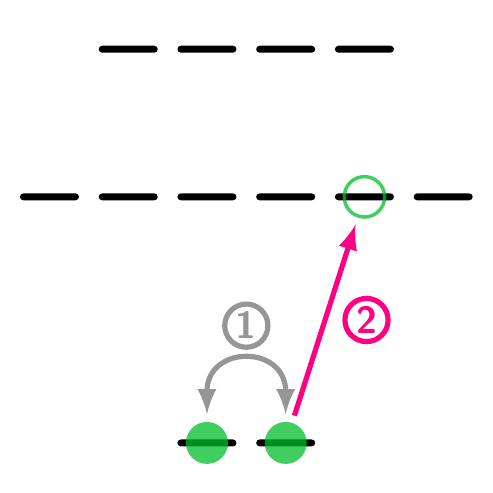}}
   \caption{Experimental scheme for photon-polarization storage. Black bars denote the Zeeman sublevels of the involved atomic levels $\SState$, $\DState$, and $\PState$. (a) Preparation: first the ion is optically pumped into the $\mState$ Zeeman sublevel of the $\SState$ ground state manifold (filled green circle); then a sequence of one radio-frequency (RF) and two optical pulses creates a coherent superposition between $\ket{\DState,m=-\frac{3}{2}}$ and $\ket{\DState,m=+\frac{3}{2}}$ (open green circles). (b) Storage: the absorption of a 854\,nm photon of arbitrary polarization (red arrows) triggers the emission of a single 393\,nm photon (wavy arrows). The detection of a $\pi$-polarized 393\,nm photon (blue) projects the atom into a superposition state in $\SState$ that corresponds to the polarization of the absorbed 854\,nm photon. (c) Read-out: electron shelving (pink arrow) and atomic-fluorescence detection distinguish the $\SState$ sublevels; depending on the desired measurement basis, another RF pulse is applied before.}
   \label{fig:scheme}
\end{figure}

\textbf{Experimental protocol and setup.} Fig.\ \ref{fig:scheme} illustrates the principle of the experiment, which adapts an earlier proposal \cite{Lloyd2001} to our atomic system, a single $\Ca$ ion \cite{Mueller2014}. A superposition state in the $\DState$ manifold is prepared as initial state for heralded absorption. Photons at 854\,nm wavelength excite this state to $\PState$. For any photon polarization, the corresponding superposition of the $\sigma^+$ and $\sigma^-$ transition is driven. Upon absorption of a photon, the ion returns to the $\SState$ ground state (with 93.5\,\% probability, given by the $\PState$ branching fractions \cite{Gerritsma2008}), releasing a single photon at 393\,nm wavelength \cite{Kurz2013}. State mapping is completed by the detection of a $\pi$-polarized 393\,nm photon; it heralds the absorption process while creating a superposition of the $\SState$ sublevels that corresponds to the polarization of the absorbed photon.

In order to verify the state transfer and quantify its fidelity, standard atomic-state analysis \cite{Haeffner2008} is performed after the detection of the heralding photon. For measuring the ion in the $\pmState$ basis of the $\SState$ manifold, the $\pState$ state is transferred (shelved) to the $\DState$ manifold. Switching on the cooling lasers then either reveals a fluorescence signal (ion in $\mState$) or not (ion in $\pState$). For projecting onto a superposition basis of $\pmState$, a resonant RF pulse effects a basis rotation before shelving.

\begin{figure}[ht]
   \includegraphics[width=0.9\columnwidth]{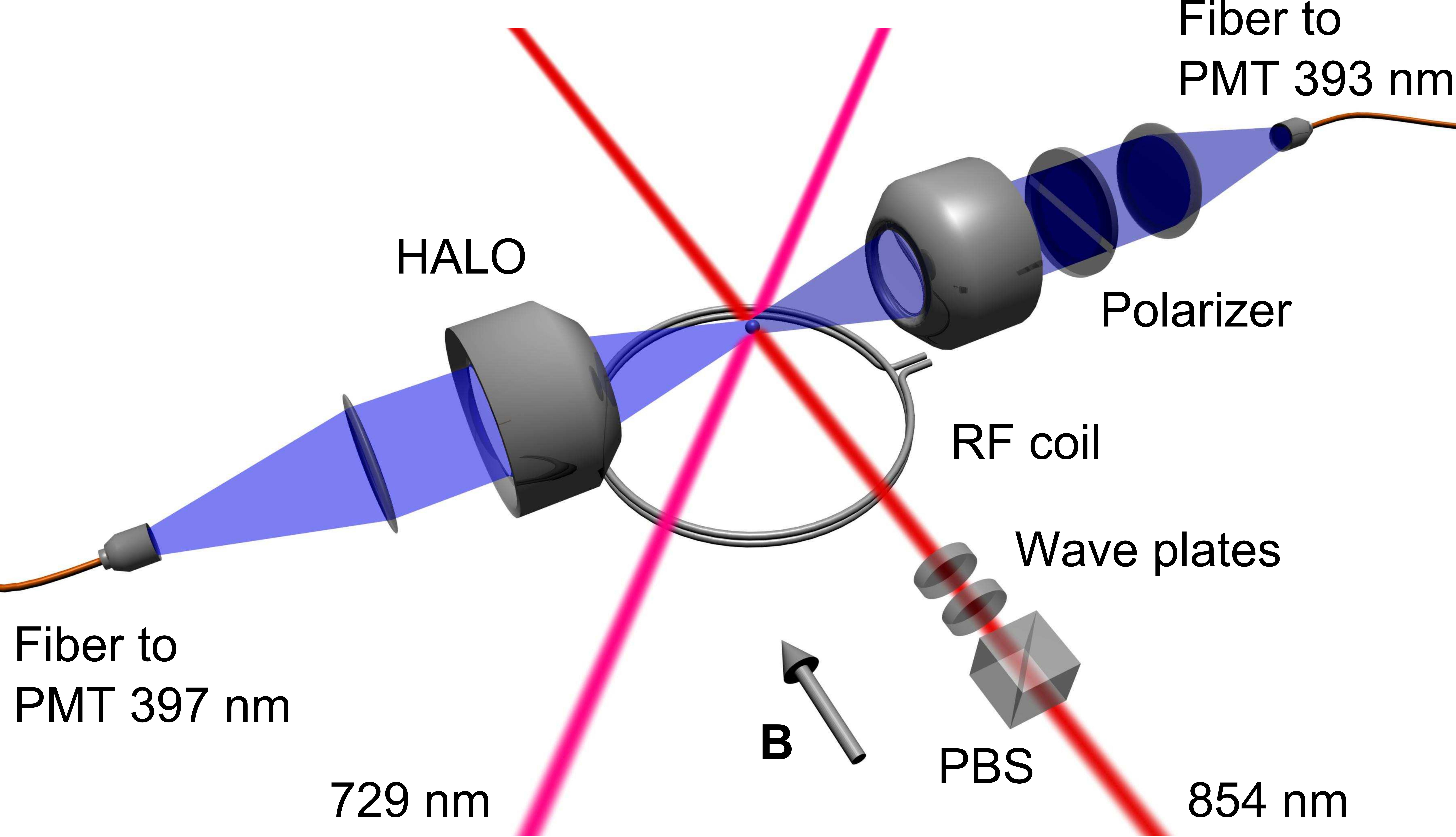}
   \caption{Experimental setup. HALO: high-NA laser objective, PMT: photomultiplier tube, PBS: polarizing beam splitter, \textbf{B}: magnetic-field direction. The ion is trapped between the HALOs.}
   \label{fig:setup}
\end{figure}

Our experimental setup is sketched in fig.\ \ref{fig:setup}. A single $\Ca$ ion is trapped in a linear radio-frequency (Paul) trap and Doppler cooled by frequency-stabilized diode lasers \cite{Rohde2010}. Cooling is facilitated by a static magnetic field, also defining the quantization axis. A narrow-band laser at 729\,nm is used for coherent manipulations on the $\SState$--$\DState$ transition. A laser at 854\,nm with variable polarization provides the photons to excite the $\DState$--$\PState$ transition. For the purpose of demonstrating our experimental protocol, we send many identical photons to the atom in each individual run; quantum state transfer from a single photon to the atom would be attained by the same procedure. In order to coherently control the magnetic-dipole transition coupling the two $\SState$ Zeeman states, we use a copper-wire coil placed below the trap. An RF current at $7.8\,\text{MHz}$ creates an oscillating magnetic field (perpendicular to the quantization axis) at the position of the ion, driving the magnetic-dipole transition. Blue photons emitted by the ion are collected through two in-vacuum high-NA laser objectives (\mbox{HALOs} \cite{Gerber2009}). Photons at 393\,nm wavelength are filtered by a polarizer such that only those scattered on $\pi$ transitions are transmitted. The photons are then coupled to optical fibers and detected by photomultiplier tubes (PMTs) whose output pulses are fed into the sequence-control unit. For single photons at 393\,nm, the pulses are in addition time-tagged and stored for later processing.

\addvspace{1em}

\begin{figure}[ht]
   \subfigure[]{\includegraphics[width=0.9\columnwidth]{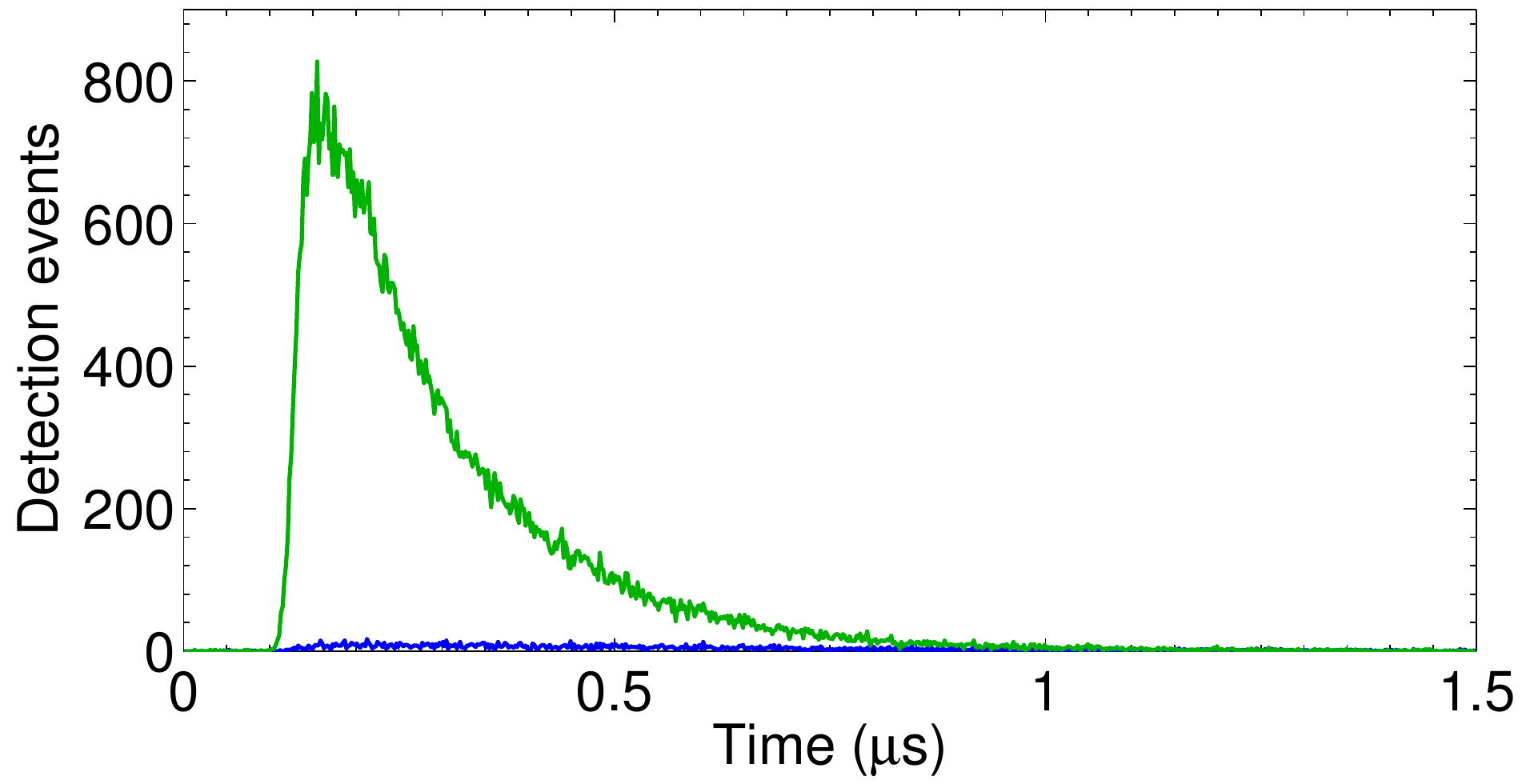} \label{fig:nonoscphoton}}
   \subfigure[]{\includegraphics[width=0.9\columnwidth]{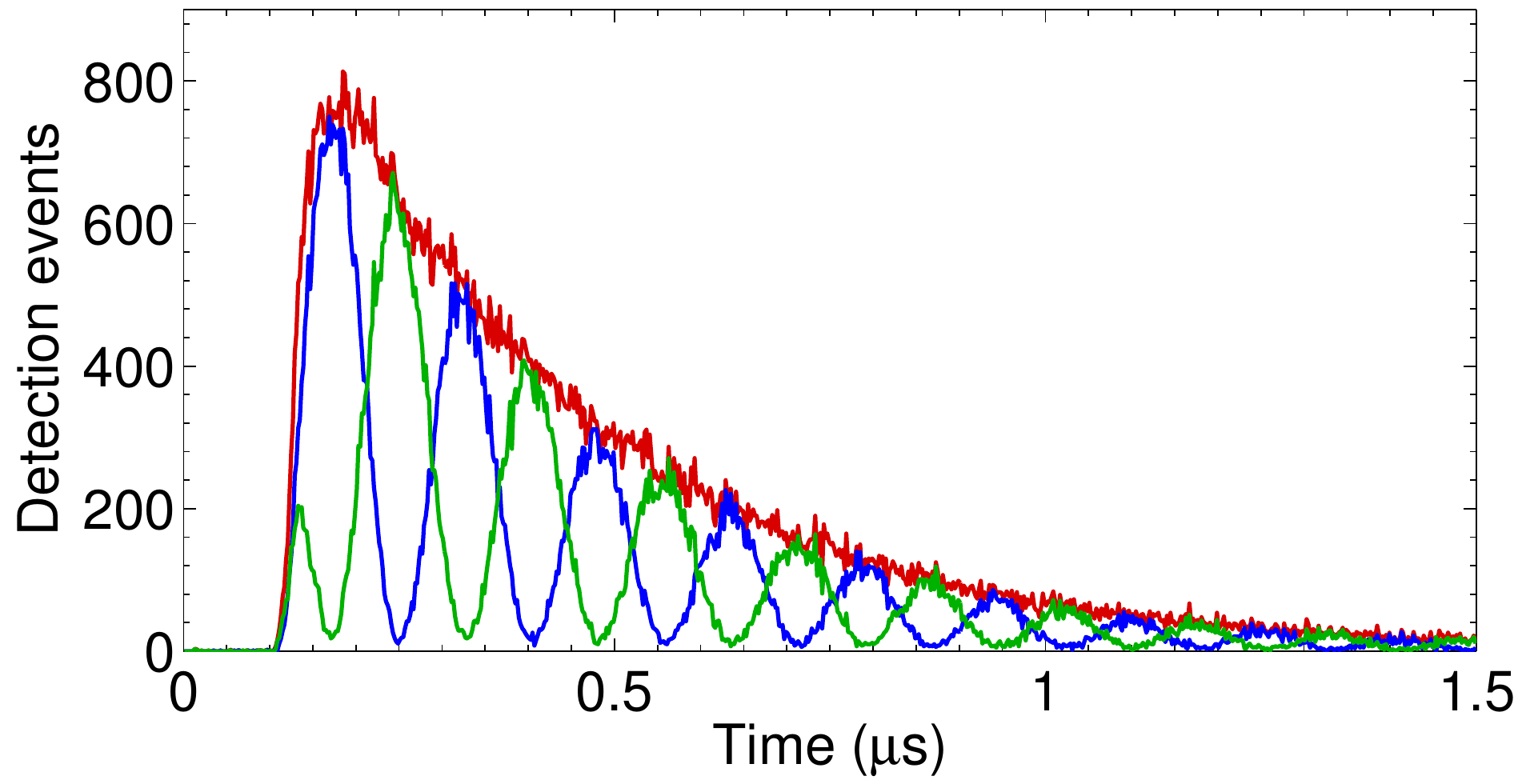} \label{fig:oscphoton}}
   \caption{Conditional arrival-time histograms of single 393\,nm photons. (a) Projection onto $\pState$ (blue) and $\mState$ (green) for right-handed circularly polarized 854\,nm light (b) Projection onto the superposition basis for linearly polarized 854\,nm light. The red curve shows the sum of the two histograms, i.e.\ the unconditional arrival-time distribution. For each graph, the bin size is 2\,ns and the overall measuring time is 20\,min.}
\end{figure}

\textbf{State transfer in circular polarization basis.} In a first experiment, we choose the 854\,nm-light polarization to be right-handed circular, driving $\sigma^+$ transitions in the atomic system. Hence, only the population in the $\ket{\DState,m=-\frac{3}{2}}$ state is excited. The detection of a single 393\,nm photon in the state $\ppi$ then ideally leaves the ion in the state $\mState$. After detecting a photon, the atomic state is measured in the $\pmState$ basis. According to the measurement outcome, the photon arrival times are sorted into two histograms. As can be seen from fig.\ \ref{fig:nonoscphoton}, the emission of photons followed by a projection of the atomic state onto $\pState$ is almost completely suppressed. As a reasonable trade-off between high quantum-state fidelity and high photon-detection efficiency, we select only detection events within the first 450\,ns of the photonic wave packet. From the number of events in each histogram within this time window, we obtain a fidelity with respect to the ideal atomic state $\mState$ of 97.8(1)\,\%. The measurement was then repeated for left-handed circularly polarized 854\,nm light, yielding a fidelity of the atomic state with respect to $\pState$ of 96.4(3)\,\%.

\addvspace{1em}

\textbf{State transfer of linear polarization.} In a second step, we change the 854\,nm-light polarization to linear. Thus a superposition of both $\sigma$ transitions is driven, with the relative phase given by the angle of the linear polarization (which is defined with respect to the plane spanned by the axis of fluorescence collection and the quantization axis). The detection of a scattered $\ppi$ photon then heralds the projection of the atomic state onto the superposition of $\pState$ and $\mState$ that corresponds to the 854\,nm-light polarization. Finally, a $\frac{\pi}{2}$ RF pulse is applied before shelving and fluorescence detection, thereby measuring the atomic state in the superposition basis. Fig.\ \ref{fig:oscphoton} shows the arrival-time histograms for the two possible measurement outcomes. The oscillations visible in the shapes of the conditional photon wave packets are explained by the difference between the Larmor frequency of the initial superposition of $\ket{\DState,m=\pm\frac{3}{2}}$ and that of the final superposition of $\ket{\SState,m=\pm\frac{1}{2}}$. Taking into account the Land\'{e} factors, the oscillation period is given by $T=\frac{h}{1.6\,\mu_B B}=160\,\text{ns}$ for a static magnetic field $B=2.8\,\text{G}$. In addition, the wave packets are approximately twice as long as for circularly polarized 854\,nm light, since linear polarizations have a relative overlap with $\sigma$ transitions of only $\frac{1}{\sqrt{2}}$. This results in lower Rabi frequencies and hence longer wave packets.

\begin{figure}[ht]
   \includegraphics[width=0.9\columnwidth]{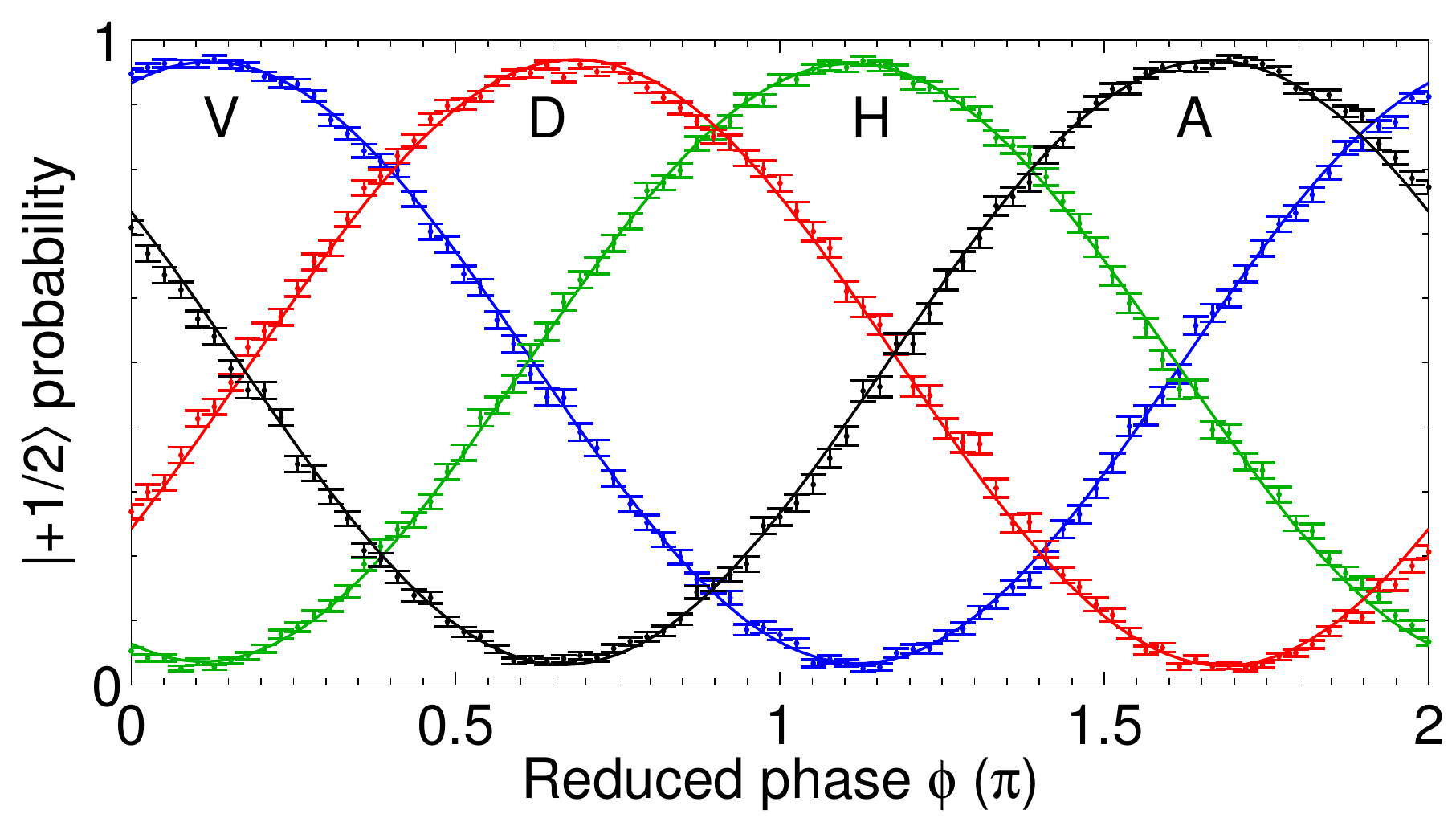}
   \caption{Probability to find the ion in state $\pState$ after the $\frac{\pi}{2}$ RF pulse as a function of the reduced phase $\phi$ for the 854\,nm-light polarizations vertical (V), horizontal (H), diagonal (D) and antidiagonal (A). The solid lines are sinusoidal fits.}
   \label{fig:fringes}
\end{figure}

In order to show that the detection of a Raman-scattered photon indeed heralds the successful transfer of the polarization state of a laser photon onto the atomic ground state, we analyze contrast and phase of the oscillations in the Raman-photon wave packets. As the phase accumulated in the $\ket{\DState,m=\pm\frac{3}{2}}$ superposition before returning to the ground state is directly given by the time duration until the photon is detected, each detection-time value $t$ converts into a phase value $\varphi=2\pi\frac{t}{T}$. As detection times differing by integer multiples of the Larmor period $T$ yield equivalent final atomic states, we use the reduced phase $\phi=\varphi\,\text{mod}\,2\pi$. For each such phase, we derive from the arrival-time histograms the probability of projecting the ion onto the state $\pState$ after the $\frac{\pi}{2}$ RF pulse. For the 450\,ns time window, these probabilities are depicted in fig.\ \ref{fig:fringes} for different linear polarizations of the 854\,nm light. We observe the expected sinusoidal behavior as a function of the reduced phase $\phi$, as well as the expected consecutive relative phase shift of $\frac{\pi}{2}$ going from V to D, H, and A polarized light. From the fitted visibility values $V$ we obtain the fidelities $F$ of the four generated atomic states with respect to the ideal states corresponding to V, H, D, and A polarization, $F=\frac{1}{2}\left(1+V\right)=96.7(1)\,\%$, 96.4(1)\,\%, 97.0(2)\,\%, and 96.9(2)\,\%, respectively. Including the previously determined fidelities for circular polarization, we arrive at an average fidelity of all six reconstructed atomic states with respect to the ideal states of 96.9(1)\,\%.

The observed quantum-state fidelities are mainly limited by two effects: Firstly, detector dark counts mimic the detection of 393\,nm photons (1.7\,\%). Secondly, spontaneous decay of population from the $\PState$ back to the $\DState$ state leaves the ion in a mixed state from which it is subsequently transferred to $\SState$; the observed contribution (1.4\,\%) is on the order of what is expected from the $\PState$ branching fractions. This limitation will not matter in experiments employing true single photons at 854\,nm since there is no second photon available for reexciting the ion. The state fidelity may also be reduced by magnetic-field noise, but here the other two effects were more relevant; in a different realization of the protocol, described below, magnetic-field noise was found to be significant.

\addvspace{1em}

\begin{figure}[ht]
   \includegraphics[width=0.9\columnwidth]{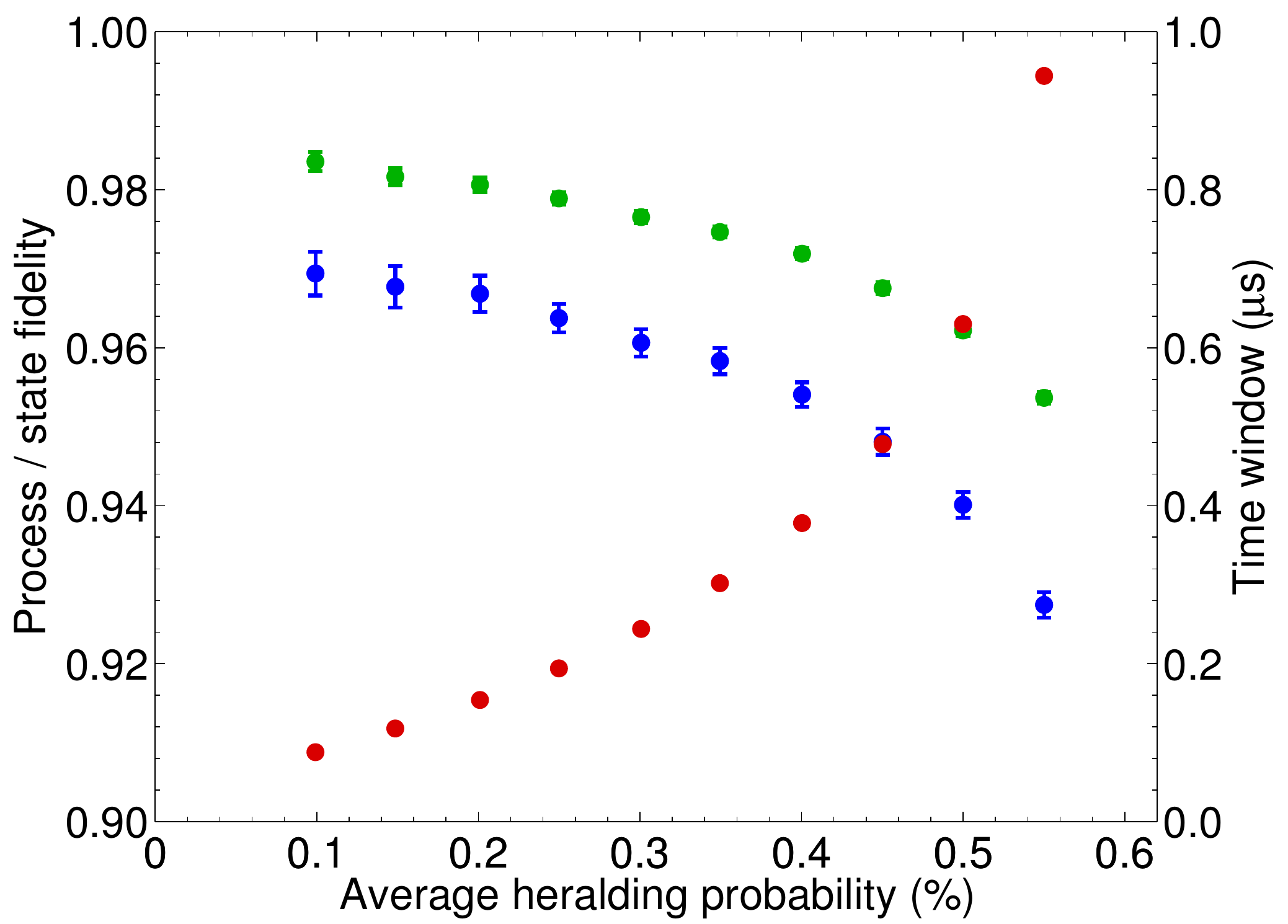}
   \caption{Process fidelity (blue dots) and average quantum-state fidelity (green dots), characterizing the mapping process, for different heralding probabilities (set through the time-window length indicated by red dots).}
   \label{fig:fidelities}
\end{figure}

\textbf{Quantum process tomography.} Another way of quantifying the transfer fidelity of the mapping process is to perform quantum process tomography. The quantum process $\varepsilon$ mapping the photon polarization state $\rho$ onto the atomic state can be expressed as $\varepsilon\left(\rho\right)=\sum_{m,n}\chi_{mn}\sigma_m\,\rho\,\sigma_n$ with the Pauli matrices $\left\{\sigma_{i=1,..,4}\right\}=\left\{\mathbbm{1},\sigma_x,\sigma_y,\sigma_z\right\}$ and the process matrix $\chi$ \cite{Chuang1997}. The value $\left|\chi_{11}\right|$ represents the identity part of the quantum process and is known as the process fidelity. It can be computed using four mutually unbiased polarization states and the resulting atomic states. For a 450\,ns time window, we derive a process fidelity of 95.0(2)\,\%, with which the atomic state reproduces the polarization of the absorbed laser photon. It is instructive to analyze the figures of merit of the transfer for different choices of the time window. Fig.\ \ref{fig:fidelities} shows the process fidelity and average state fidelity as a function of the photon-detection probability, the latter being set by the choice of the detection-time window. For our 450\,ns time window, we achieve 0.438(1)\,\% average detection efficiency.

\subsection*{Discussion}

Reference \cite{Mueller2014} discusses various schemes for mapping the polarization state of an absorbed 854\,nm photon onto the atomic ground-state qubit, of which we also implemented an alternative one: Instead of preparing the ion in a coherent superposition of the $\ket{\DState,m=\pm\frac{3}{2}}$ states for absorption, we create a superposition of the states $\ket{\DState,m=\pm\frac{5}{2}}$. The 854\,nm light then transfers the ion to the $\SState$ ground state by emitting a 393\,nm photon on either the $\sigma^+$ or $\sigma^-$ transition. For linearly polarized 854\,nm light, the polarization of the scattered photon is maximally entangled with the final atomic spin state: $\ket{\Psi_{\text{ion--photon}}}=\frac{1}{\sqrt{2}}\left(\pState\pSigma+{\mathrm e}^{i\phi}\mState\mSigma\right)$. As emitted photons are collected perpendicularly to the quantization axis, only one specific superposition of $\pSigma$ and $\mSigma$ is observed (after removing the polarizer). The detection of the scattered photon thus projects the atomic state onto the superposition of $\pState$ and $\mState$ that corresponds to the 854\,nm-light polarization. For the 450\,ns detection-time window, we achieve 91.7(5)\,\% process fidelity, 94.8(2)\,\% average state fidelity, and 0.318(1)\,\% heralding probability. The reduced fidelity as compared to the previous scheme is mostly attributed to the non-linearly \cite{Mount2013} higher sensitivity of the $\ket{\pm\frac{5}{2}}$ states to magnetic-field noise.

In conclusion, we demonstrated heralded, high-fidelity mapping of a photonic polarization state onto an atomic quantum bit. After preparing a single $\Ca$ ion in a coherent superposition state, the absorption of a photon on the $\DState$--$\PState$ transition maps the photonic polarization state onto the atomic $\SState$ Zeeman qubit. Successful mapping is heralded by the detection of a Raman-scattered photon at 393\,nm wavelength, without perturbing the atomic qubit state. By virtue of our high-NA light-collection optics, the heralding probability amounts to 0.438(1)\,\%. The process fidelity is 95.0(2)\,\% and the average fidelity of the resulting atomic state with respect to the initial polarization amounts to 96.9(1)\,\%. While in this demonstration we used laser photons to obtain a high interaction rate, the scheme can readily be applied to single photons, for example from a resonant SPDC source \cite{Haase2009}. Upon absorption of one photon, detection of a 393\,nm Raman photon will then herald the entanglement of the partner photon with the ion. This scheme shall be extended to having both entangled partner photons absorbed by ions in remote traps \cite{Sangouard2013}. Coincidental detection of two 393\,nm photons then heralds the entanglement of the two spatially separated ions and completes photon-to-atom entanglement transfer.

\subsection*{Methods}

\textbf{Experimental sequence.} All laser beams are controlled through acousto-optic modulators (AOMs). The sequence starts with a $5\,\mu\text{s}$ pump pulse of $\sigma^-$ polarized light at 397\,nm. This prepares the ion in the $\mState$ state of the $\SState$ manifold with 99.82(1)\,\% probability. Then a coherent superposition of the $\pmState$ states is created by an RF pulse of $2.8\,\mu\text{s}$ duration. Subsequent 729\,nm pulses, each lasting $9.6\,\mu\text{s}$, transfer the superposition to the $\ket{\pm\frac{3}{2}}$ states of the $\DState$ manifold. The 854\,nm beam from which a single photon is absorbed is switched on for $3\,\mu\text{s}$ at $12\,\mu\text{W}$ optical power, yielding $\sim 9\,\text{MHz}$ Rabi frequency for a $125\,\mu\text{m}$ beam waist. If the Raman-scattered photon is not detected (in most cases), we apply Doppler cooling for $10\,\mu\text{s}$. Including delays in the control electronics, we achieve a repetition rate of 18\,kHz.

\textbf{Fluorescence detection.} Light emitted by the ion is collected through two high-NA laser objectives (HALOs, $\text{NA}=0.4$), each covering $\sim 4.2\,\%$ of the total solid angle, and coupled to multi-mode optical fibers with 83\,\% efficiency. The light is then detected by photomultiplier tubes with $\sim 28\,\%$ quantum efficiency. For isotropic emission, this yields a total detection efficiency of 0.98\,\%.

\textbf{Atomic-state analysis.} After detecting the Raman photon at 393\,nm, the atomic state is analyzed: depending on whether a superposition basis is used, first a rotation in the $\SState$ manifold is performed by another RF pulse. Then we apply a shelving pulse at 729\,nm, which transfers the $\pState$ state to the $\DState$ manifold, before switching on the cooling lasers. A projection onto the $\SState$ state is signaled by the onset of resonance fluorescence, detected at a rate of $1.07\cdot 10^5\,\text{s}^{-1}$. A projection onto $\DState$ leaves the ion dark, resulting in a detection rate of $60\,\text{s}^{-1}$ from detector dark counts and laser stray light. Integrating the detection rate for $100\,\mu\text{s}$ allows us to discriminate between the two atomic states with 99.986\,\% fidelity.

\textbf{Indistinguishability of 393\,nm photons.} It is important to note that in a single-photon realization of the protocol, the arrival time distribution of the Raman-scattered photon at 393\,nm will not contain any information about the polarization of the single absorbed 854\,nm photon. With 854\,nm laser light we observe different distributions for different polarizations, as in figs.\ \ref{fig:nonoscphoton} and \ref{fig:oscphoton}. For a single-photon realization, however, the relevant quantity is the peak value of this measured distribution, which is identical for the two cases; the subsequent shape only reflects the temporal behavior of the $\DState$-state population for the case of laser excitation.

\bibliography{bibliography}

\subsection*{Acknowledgments}

We acknowledge support by the BMBF (QuOReP project, QSCALE Chist-ERA project) and the German Scholars Organization / Alfried Krupp von Bohlen und Halbach-Stiftung.

\subsection*{Author contributions}

C.K., M.S.\ and P.E.\ contributed equally to this work; J.H.\ set up the narrow-band laser; C.K., M.S.\ and P.E.\ prepared the experiment and acquired and analyzed the data; P.M.\ performed numerical calculations; J.E.\ planned and supervised the project; C.K.\ and J.E.\ wrote the manuscript.

The authors declare no competing financial interests.

\end{document}